\documentclass[twoside]{article}
\usepackage{fleqn,espcrc2}

\usepackage{graphicx}
\usepackage[figuresright]{rotating}

\newcommand{\ua}{\uparrow}
\newcommand{\da}{\downarrow}

\newcommand{\AmS}{{\protect\the\textfont2
  A\kern-.1667em\lower.5ex\hbox{M}\kern-.125emS}}

\title{Probabilistic representation of fermionic lattice systems}

\author{
Matteo Beccaria\address{
Dipartimento di Fisica, Universit\`a di Lecce, Via Arnesano,
73100 Lecce, Italy}, 
Carlo Presilla\address{
Dipartimento di Fisica, Universit\`a di Roma ``La Sapienza'',
Piazzale A. Moro 2, 00185 Roma, Italy \\
and INFM Unit\`a di Ricerca di Roma ``La Sapienza''},
Gian Fabrizio De Angelis\address{
Dipartimento di Fisica, Universit\`a di Lecce, Via Arnesano,
73100 Lecce, Italy}
and Giovanni Jona-Lasinio\address{
Dipartimento di Fisica, Universit\`a di Roma ``La Sapienza'',
Piazzale A. Moro 2, 00185 Roma, Italy}}

\begin{document}

\begin{abstract}
We describe an exact Feynman-Kac type formula
to represent the dynamics of fermionic lattice systems.
In this approach the real time or Euclidean time dynamics is 
expressed in terms of the stochastic evolution of a collection 
of Poisson processes.
From this formula we derive a family of algorithms for Monte Carlo 
simulations, parametrized by the jump rates of the Poisson processes.
\end{abstract}

\maketitle

Quantum Monte Carlo methods are powerful techniques for the numerical 
evaluation of the properties of quantum lattice systems.
In the case of fermion systems \cite{TC,LINDEN,ZCG,CS} there are 
special features connected with the anticommutativity of the variables 
involved.
In a recent paper \cite{Expanded} progress has been made
by providing an exact probabilistic representation for the dynamics of a 
Hubbard model.
Here we illustrate the basic formula 
while for details we refer to \cite{Expanded}.

Let us consider the Hubbard Hamiltonian
\begin{eqnarray}
\label{Hubbard}
H &=& - \sum_{i=1}^{|\Lambda|} \sum_{j=i+1}^{|\Lambda|} 
\sum_{\sigma=\ua\da} \eta_{ij}
(c^\dag_{i\sigma} c^{}_{j\sigma} + c^\dag_{j\sigma} c^{}_{i\sigma}) 
\nonumber \\&&+ 
\sum_{i=1}^{|\Lambda|} \gamma_i~ 
c^\dag_{i\ua} c^{}_{i\ua}~c^\dag_{i\da} c^{}_{i\da},
\end{eqnarray}
where $\Lambda\subset Z^d$ is a finite $d$-dimensional lattice
with cardinality $|\Lambda|$,
$\{1, \dots,|\Lambda|\}$ some total ordering of the lattice points,
and $c_{i \sigma}$  the usual anticommuting destruction operators
at site $i$ and spin index $\sigma$.
In this paper, we are interested in evaluating the matrix elements
$\langle {\bf n}' |e^{- Ht} | {\bf n} \rangle$
where ${\bf n}= (n_{1 \ua},n_{1 \da}, \ldots, n_{|\Lambda| \ua},
n_{|\Lambda| \da})$ are the occupation numbers taking the values 0 or 1.
The total number of fermions per spin component is a conserved
quantity, therefore we consider only configurations ${\bf n}$ and 
${\bf n}'$ such that 
$\sum_{i=1}^{|\Lambda|} n'_{i\sigma} = \sum_{i=1}^{|\Lambda|} n_{i\sigma}$
for $\sigma = \ua \da$.
In the following we shall use the mod 2 addition
$n \oplus n'= (n+n') \bmod 2$.

Let $\Gamma =\{(i,j), 1\le i< j\le |\Lambda|: \eta_{ij}\neq 0\}$ 
and $|\Gamma|$ its cardinality.
For simplicity, we will assume that $\eta_{ij}=\eta$ 
if $(i,j) \in \Gamma$ and $\gamma_i=\gamma$.
By introducing
\begin{eqnarray}
\label{lambda}
\lambda_{ij \sigma}({\bf n}) &\equiv& 
\langle {\bf n} \oplus {\bf 1}_{i\sigma} \oplus {\bf 1}_{j\sigma}|
c^\dag_{i\sigma} c^{}_{j\sigma} + c^\dag_{j\sigma} c^{}_{i\sigma}
|{\bf n}\rangle 
\nonumber \\ &=&
(-1)^{n_{i\sigma} + \cdots + n_{j-1 \sigma}} 
\left[ n_{j \sigma} (n_{i \sigma} \oplus 1) 
\right. \nonumber \\&& \left.-
n_{i \sigma} (n_{j \sigma} \oplus 1) \right],  
\end{eqnarray}
where ${\bf 1}_{i\sigma}=(0,\ldots,0,1_{i\sigma},0,\ldots,0)$, and 
\begin{eqnarray}
V({\bf n}) &\equiv& 
\langle {\bf n}|H|{\bf n}\rangle =
\gamma \sum_{i=1}^{|\Lambda|} n_{i \ua} n_{i \da}, 
\label{potential}
\end{eqnarray}
the following representation holds
\begin{eqnarray}
\label{TheFormulaa}
\langle {\bf n}'|e^{-Ht} | {\bf n}\rangle &=&  {\bf E} \left(
\delta_{ {\bf n}' , {\bf n} \oplus {\bf N}^t } 
{\cal M}^t \right) 
\end{eqnarray}
\begin{eqnarray}
\label{TheFormulab}
{\cal M}^t &=& \exp \biggl\{ 
\sum_{(i,j)\in\Gamma} \sum_{\sigma=\ua\da} 
\int_{[0,t)} \!\!\!\!\!\!
\log \left[ \eta \rho^{-1} 
\right. \nonumber \\ &&\times \left.
\lambda_{ij \sigma} ({\bf n} \oplus {\bf N}^s) \right] dN^s_{ij\sigma} 
\nonumber \\ && 
- \int_0^t V({\bf n} \oplus {\bf N}^{s}) ds 
+ 2|\Gamma|\rho t \biggr\}.
\end{eqnarray}
Here, $\{N_{ij\sigma}^t\}$, $(i,j) \in \Gamma$,  is a family of 
$2|\Gamma|$ independent Poisson processes with parameter $\rho$ and 
${\bf N}^t= (N_{1 \ua}^t,N_{1 \da}^t, \ldots, N_{|\Lambda| \ua}^t,
N_{|\Lambda| \da}^t)$ are $2|\Lambda|$ stochastic processes defined as
\begin{equation}
N^t_{k \sigma} = \sum_{
(i,j) \in \Gamma:~i=k~{\rm or}~j=k }
N^t_{ij \sigma}.
\end{equation}
We remind that a Poisson process $N^t$ with parameter $\rho$ is a jump 
process characterized by the following probabilities:
\begin{equation}
P\left( N^{t+s} - N^t=k \right) = {(\rho s)^k \over k !} e^{-\rho s}.
\label{Pp}
\end{equation}
Its trajectories are piecewise-constant increasing integer-valued 
functions continuous from the left.
The stochastic integral $\int dN^t$ is just an ordinary 
Stieltjes integral
\begin{eqnarray}
\int_{[0,t)} f(s,N^s) dN^s = \sum_{k:~ s_k <t} f(s_k,N^{s_k}),
\nonumber
\end{eqnarray}
where $s_k$ are random jump times having probability density
$p(s)=\rho e^{- \rho s}$.
Finally, the symbol ${\bf E}( \ldots )$ is the expectation of the 
stochastic functional within braces.
We emphasize that a similar representation holds for the real
time matrix elements $\langle {\bf n}' |e^{- iHt} | {\bf n} \rangle$.

Summarizing, we associate to each $\eta_{ij} \neq 0$ 
a link connecting the sites $i$ and $j$ and assign to it a pair of 
Poisson processes $N^t_{ij \sigma}$ with $\sigma=\ua\da$. 
Then, we assign to each site $i$ and spin component $\sigma$ 
a stochastic process $N^t_{i \sigma}$ which is the sum of all 
the processes associated with the links incoming at that site
and having the same spin component.
A jump in the link process $N^t_{ij \sigma}$ implies a jump in
both the site processes $N^t_{i \sigma}$ and $N^t_{j \sigma}$. 
Equations (\ref{TheFormulaa}) and (\ref{TheFormulab}) are immediately
generalizable to non identical parameters $\eta_{ij}$ and $\gamma_i$. 
In this case, it may be convenient to use Poisson processes 
$N^t_{ij \sigma}$ with different parameters $\rho_{ij\sigma}$. 

In order to construct an efficient algorithm for evaluating 
(\ref{TheFormulaa}-\ref{TheFormulab}), 
we start by observing that the functions 
$\lambda_{ij\sigma}({\bf n} \oplus {\bf N}^s)$ vanish 
when the occupation numbers $n_{i\sigma} \oplus N_{i\sigma}^s$ and 
$n_{j\sigma} \oplus N_{j\sigma}^s$ are equal.
We say that for a given value of $\sigma$ the link $ij$ is active
at time $s$ if $\lambda_{ij\sigma}({\bf n} \oplus {\bf N}^s)\neq 0$.
We shall see in a moment that only active links are relevant. 
Let us consider how the stochastic integral in (\ref{TheFormulab}) builds
up along a trajectory defined by considering the time ordered 
succession of jumps in the family $\{ N^t_{ij\sigma} \}$.
The contribution to the stochastic integral in the 
exponent of (\ref{TheFormulab}) at the first jump time of a link, 
for definiteness suppose that the link $i_1j_1$ with spin component
$\sigma_1$ jumps first at time $s_1$, is
\begin{eqnarray}
\log \left[ \eta \rho^{-1} \lambda_{i_1j_1 \sigma_1} 
({\bf n} \oplus {\bf N}^{s_1}) \right] ~\theta(t-s_1),
\nonumber
\end{eqnarray}
where ${\bf N}^{s_1}={\bf 0}$ due the assumed left continuity.
Therefore, if the link $i_1j_1\sigma_1$ was active at time 0 we obtain 
a finite
contribution to the stochastic integral otherwise we obtain $-\infty$.
If $s_1 \geq t$ we have no contribution to the stochastic 
integral from this trajectory. 
If $s_1 < t$ a second jump of a link, suppose $i_2j_2$ with spin
component $\sigma_2$, can take place
at time $s_2>s_1$ and we obtain a contribution 
\begin{eqnarray}
\log \left[ \eta \rho^{-1} \lambda_{i_2j_2 \sigma_2} 
({\bf n} \oplus {\bf N}^{s_2}) \right] ~\theta(t-s_2).
\nonumber
\end{eqnarray}
The analysis can be repeated by considering an arbitrary number of jumps.
Of course, when the stochastic integral is $- \infty$, which is the 
case when some $\lambda=0$, there is no contribution to the expectation. 
The other integral in (\ref{TheFormulab}) is an ordinary integral of
a piecewise constant bounded function. 

We now describe the algorithm. 
The only trajectories to be 
considered are those associated to jumps of the active links.
This guarantees conservation of the total number of fermions per spin component.
The active links can be determined after each jump by inspecting the
occupation numbers of the sites in the set $\Gamma$ according to the 
rule that the link $ij$ is active for the spin component $\sigma$ if 
$n_{i\sigma} + n_{j \sigma} =1$. 
We start by determining the active links in the initial configuration 
${\bf n}$ assigned at time $0$ and make an extraction with uniform 
distribution to decide which of them jumps first. 
We then extract the jump time $s_1$ according to the probability
density $p_{A_1}(s)=A_1 \rho \exp(- A_1 \rho s)$ where $A_1$ 
is the number of active links before the first jump takes place.
The contribution to ${\cal M}^t$ at the time of the first jump 
is therefore, up to the factor 
$\exp \left( -2|\Gamma|\rho t \right)$,
\begin{eqnarray}
&& \eta \rho^{-1} \lambda_{i_1j_1 \sigma_1} ({\bf n} \oplus 
{\bf N}^{s_1})
e^{ -V({\bf n} \oplus {\bf N}^{s_1}) s_1 }
\nonumber \\ &&\times
e^{-(2|\Gamma|-A_{1}) \rho s_1} \theta(t-s_1)
\nonumber \\ && 
+ e^{ -V({\bf n} \oplus {\bf N}^{t}) t }
e^{-(2|\Gamma|-A_{1})\rho t} ~\theta(s_1-t),  
\nonumber
\end{eqnarray}
where $\exp [-(2|\Gamma|-A_{1})\rho s]$ is the probability 
that the $2|\Gamma|-A_{1}$ non active links do not jump in the 
time interval $s$.
The contribution of a given trajectory is obtained by multiplying the
factors corresponding to the different jumps until the last jump 
takes place later than $t$.
For a given trajectory we thus have 
\begin{eqnarray}
{\cal M}^t &=& \prod_{k \geq 1} \Bigl [
\eta \rho^{-1} \lambda_{i_kj_k \sigma_k} ({\bf n} \oplus {\bf N}^{s_k})
\nonumber \\ &&\times
e^{ [A_{k} \rho -V({\bf n} \oplus {\bf N}^{s_k})] (s_k-s_{k-1}) }
~\theta(t-s_k) 
\nonumber \\ && +
e^{ [A_{k} \rho -V({\bf n} \oplus {\bf N}^{t})] (t-s_{k-1}) }
~\theta(s_k-t) \Bigr].
\label{Mt}
\end{eqnarray}
Here, $A_k=A({\bf n} \oplus {\bf N}^{s_k})$ is the number of active 
links in the interval $(s_{k-1}, s_k]$ and $s_0=0$.
Note that the exponentially increasing factor 
$\exp \left( 2|\Gamma|\rho t \right)$ in (\ref{TheFormulab}) cancels
out in the final expression of ${\cal M}^t$.
The analogous expression of ${\cal M}^t$ for real times is simply 
obtained by replacing $\eta \to i \eta$ and $\gamma \to i \gamma$.
The algorithm can be improved by the usual methods of reconfiguration
and importance sampling.

In principle, the algorithms parametrized by $\rho$ are all equivalent
as (\ref{TheFormulaa}-\ref{TheFormulab}) holds for any choice of
the Poisson rates.
However, since we estimate the expectation values with a finite
number of trajectories, this may introduce a systematic error.
It can be shown that the best performance is obtained for the 
natural choice $\rho \sim \eta$ 
independently of the interaction strength.
In this case our algorithm coincides with the Green function Monte Carlo
method in the limit when the latter becomes exact 
as discussed in \cite{Expanded}.

\vfill

\end{document}